\documentclass[showpacs,aps,amsmath,amssymb,twocolumn]{revtex4}
\usepackage[latin1]{inputenc}
\usepackage{dcolumn}
\usepackage{bm}
\usepackage{verbatim} 
\usepackage[]{graphicx}			
\usepackage{color}			
\newcommand{\be}{\begin{equation}}
\newcommand{\ee}{\end{equation}}
\newcommand{\ben}{\begin{eqnarray}}
\newcommand{\een}{\end{eqnarray}}
\newcommand{\bes}{\begin{subequations}}
\newcommand{\ees}{\end{subequations}}
\newcommand{\bb}{\bibitem}

\newcommand{\bfi}{\begin{figure}}
\newcommand{\efi}{\end{figure}}
\newcommand{\bc}{\begin{center}}
\newcommand{\ec}{\end{center}}

\begin{document}
\title{New Results on Compact Structures}

\author{D. Bazeia,$^{1,2}$ L. Losano,$^{1,2}$ and R. Menezes$^{2,3}$}

\affiliation{$^1$Departamento de F\'\i sica, Universidade Federal da Para\'\i ba, 58051-970 Jo\~ao Pessoa, PB, Brazil\\ $^2$Departamento de F\'\i sica, Universidade Federal de Campina Grande, 58109-970 Campina Grande, PB, Brazil and \\ $^3$Departamento de Ci\^encias Exatas, Universidade Federal da Para\'\i ba, 58297-000 Rio Tinto, PB, Brazil.}

\date{\today}

\begin{abstract}
We investigate the presence of localized solutions in models described by a single real scalar field with generalized dynamics. The study offers a method to solve very intricate nonlinear ordinary differential equations, and we illustrate the results with some examples on localized structures with compact profile, in models with polynomial and nonpolynomial interactions. We also show that the compact solutions we have found are all linearly stable.
\end{abstract}


\maketitle


{\it Introduction.}  A remarkable phenomenon that occurs in nonlinear science is the existence of solitons, which spring from the interplay between dispersion and nonlinearity \cite{soliton}. In the presence of nonlinear dispersion, however, solitons may acquire spatial profiles with compact support \cite{compact1}. They have been investigated in several different contexts \cite{compact1,compact2,css} and, in one space dimension, the compact excitations are spacelike structures similar to kinks, which appear in relativistic systems in $(1,1)$ spacetime dimensions \cite{book}. This identification has led us to investigate compactons in relativistic scalar field theories, a subject much less explored then its nonrelativistic counterpart \cite{compact1,compact2,css}.  Although we deal with relativistic systems, the search for static solutions leads to differential equations that may be used to map nonrelativistic systems, so it is not hard to navigate from relativistic to nonrelativistic systems in the current context. For instance,  studies on compact traveling waves in systems investigated in Ref.~\cite{css} lead us to first-order differential equations, very much similar to the first-order equations that we investigate in the current work.

Localized static structures such as kinks are of great importance to study issues of current interest in high energy physics \cite{book}. They only require scalar fields in one space dimension, and have gained a lot of attention in the last years, because they map interesting phenomena in physics \cite{kink} and contribute to model braneworld scenarios with a single extra dimension of infinite extent \cite{RS,brane1}. Kinks have energy densities that vanish asymptotically. Compactons are different, since their energy densities vanish outside a compact space. However, they also appear in distinct scenarios in nonlinear science \cite{compact1,compact2,css,defect}. Furthermore, the difficulty in implementing nonlinear dispersion is being circumvented due to management techniques for soliton control. For instance, in Ref.~\cite{optics} one finds several possibilities to adjust parameters to control nonlinear features of continuun and periodic systems. 

In this work we offer a simple and direct way to solve very intricate problems engendering nonlinearity and nonlinear dispersion, described by ordinary differential equations of current interest to physics. The method springs from Refs. \cite{first,blm}, motivated by the first-order framework \cite{first}, of interest to investigate generalized models \cite{kf,twin}, and the procedure of Ref.~\cite{blm}, which has been used in a diversity of contexts in Ref.~\cite{deform}. Here we focus mainly on solutions with compact profile, and we show how to construct new compact structures in models described by a single real scalar field in $(1,1)$ spacetime dimensions. The approach is robust, and we also use it to construct distinct models, supporting the very same static structure.

{\it The procedure.} Let us start writing the Lagrange density for the generalized model. We work with a single real scalar field $\phi=\phi(x,t)$ in $(1,1)$ spacetime dimensions, and we use dimensionless field, space and time coordinates, and coupling constants, for simplicity. The metric is $(+,-)$. In the case of a standard model, the Lagrange density has the form
\be\label{sm}
{\cal L}(\phi,\partial_\mu\phi)=\frac12\partial_\mu\phi\partial^\mu\phi-V(\phi).
\ee
We make the investigation more general introducing
\be\label{gl}
{\cal L}(\phi,X)=F(X)-V(\phi),
\ee
where $X=\partial_\mu\phi\partial^\mu\phi/2.$ 
In the generalized Lagrange density \eqref{gl}, the dependence on $\phi$ and $X$ may be set at will, under the guidance of positivity of energy.

The above Lagrange density allows obtaining the energy-momentum tensor
\be\label{tensor}
T_{\mu\nu}={\cal L}_{X}  \partial_{\mu} \phi \partial_{\nu} \phi- g_{\mu\nu} {\cal L}\,,
\ee
where ${\cal L}_{X}=\partial{\cal L}/\partial_X$. Also, the equation of motion for the scalar field $\phi=\phi(x,t)$ has the form
\be\label{eqm1}
 \partial_{\mu} \left({\cal L}_{X}  \partial^{\mu} \phi\right)={\cal L}_{\phi}\,.
\ee
If  we search for static solution, $\phi=\phi(x)$, we then get
\be\label{eqm2}
({\cal L}_{X}+2 {\cal L}_{X X} X) \phi^{\prime\prime} +{\cal L}_{\phi}=0\,,
\ee
where prime denotes derivative with respect to $x$ and now $X=-\phi^{\prime\,2}/2$.  
This equation can be integrated to give
\be\label{stress}
{\cal L}-2 {\cal L}_{X}X=0.
\ee
We  emphasize that solutions to the above equation obey the stressless condition $\tau(x)=T_{11}={\cal L}+{\cal L}_X\phi^{\prime2}=0$.
Moreover, the energy density for static solutions can be written as $\rho(x)=T_{00} =-{\cal L}(\phi(x),\phi^\prime(x))$ and so from the stressless condition we have
\be
\rho(x)= {\cal L}_{X}\phi^{\prime\,2}\,.
\ee 

We follow \cite{first} and we introduce another function, $W=W(\phi)$, such that
\be\label{Lw}
{\cal L}_{X}\phi^\prime=W_{\phi}\,.
\ee
Although $W(\phi)$ is in principle a general function of $\phi$, the above equation has to be evaluated with $\phi=\phi(x)$, as a static configuration. It allows to write $\rho(x)=dW/dx$ and the energy becomes $E=\Delta W=W(\phi(\infty))-W(\phi(-\infty))$.
We use the equation of motion \eqref{eqm2} and \eqref{Lw} to write
\be\label{foe}
W_{\phi\phi}\,\phi^{\prime}=-{\cal L}_{\phi}\,,
\ee
which is a first-order equation for the field configuration. We note that solutions to this first-order equation \eqref{foe} solve the equation of motion \eqref{eqm2}. We emphasize that, since we are changing $X\to F(X)$ to get to the generalized model, and since $F(X)$ is nonlinear, one has to check explicitly that solutions to the
first-order \eqref{foe} solve the equation of motion \eqref{eqm2}, for every specific problem under investigation.

We focus attention on the first-order Eq.~\eqref{foe}, which can be written as $\phi^\prime=V_\phi/W_{\phi\phi}$, showing $\phi^\prime$ is a function of the field $\phi$ itself. For simplicity, we write $\phi^\prime\!=\!V_\phi/W_{\phi\phi}\!=\!R(\phi)$. We see from Eq.~\eqref{Lw} that $R(\phi)\!=\!W_\phi(\phi)$, if we consider the standard model \eqref{sm}.
This result shows that $R(\phi)$ generalizes the superpotential $W(\phi)$ which appears in the bosonic sector of a the supersymmetric theory, with standard kinematics. This issue is interesting, and may help us to construct supersymmetric extension of the generalized model, a problem to be considered elsewhere.

We now see from $\phi^\prime=R(\phi)$ and Eq.~\eqref{stress}, that it is always possible to write $V(\phi)$ as $V(R(\phi))$.
This allows us to use the procedure of Ref.~\cite{blm} to propose and solve new generalized models described by a real scalar field in $(1,1)$
spacetime dimensions. To make the argument explicit, we consider the previous model \eqref{gl} and another one, described by the scalar field $\chi(x,t)$, with similar Lagrange density
\ben\label{model2}
\tilde{\cal L}(\chi,Y)&=&G(Y)- U(\chi)\,,\,\,\,Y=\frac12 \partial_\mu \chi \partial^\mu \chi\,,
\een
where $G(Y)$ and $U(\chi)$ identify the new model. We suppose that the two models behave adequately, and that they support nontrivial static solutions. In the second model \eqref{model2}, we write $G(Y)$ explicitly, but we leave $U(\chi)$ arbitrary, to be constructed as follows: We employ the first-order formalism and write $\phi^\prime=R(\phi)$, and $\chi^\prime=S(\chi)$, where $S(\chi)$ is a function of $\chi$, similar to $R(\phi)$. $S(\chi)$ is not known yet, because we do not know $U(\chi)$. We start from the model \eqref{gl}, with known $F(X)$ and $V(\phi)$, and with a known static solution $\phi(x)$. We introduce another function, $g=g(\chi)$, differentiable, and in $R(\phi)$ we now change $\phi\to g(\chi)$ such that $R(\phi\to g(\chi))/f^\prime(\chi)$ is now equal to
$S(\chi)$; that is, we define
\be\label{S} 
S(\chi)\equiv \frac{{R}(\phi \to g(\chi))}{g^{\prime}(\chi)}.
\ee
With $G(Y)$ and $S(\chi)$ we then use ${\tilde{\cal L}}-2 {\tilde{ \cal L}}_{Y}Y=0,$
which is similar to Eq.~\eqref{stress}, to construct the potential $U(\chi)$. The procedure implies that if $\phi(x)$ is solution of the model ${\cal L}(\phi, X)$, then $\chi(x)=g^{-1}(\phi(x))$ is solution of the new model $\tilde{\cal L}(\chi, Y)$, with $g^{-1}(\phi)$ standing for the inverse of $g(\chi)$. It allows the construction of the new model and the corresponding static solution, altogether.

It is interesting to note that the procedure of changing $R(\phi)\to S(\chi)\equiv R(\phi\to g(\chi))/g^{\prime}(\chi)$ only modifies the potential. However, we see from ${{\cal L}}-2 {\cal L}_{X}X=0$ and from ${\tilde{\cal L}}-2 {\tilde{ \cal L}}_{Y}Y=0,$ how $F(X)$ and $G(Y)$ play the game, contributing to the respective models and solutions.

{\it Illustrations.} To understand how the procedure works explicitly, we note that the simplest deformation, $g(\chi)=\chi$, which is the identity function, furnishes models with the very same static solution.
We consider, for instance, the standard model \eqref{sm} with the potential
$V(\phi)=\frac12(1-\phi^2)^2,$
and the generalized model
\be
\tilde{\cal L}=(-1)^{n-1}\frac{2^{n-1}}{n}Y^{n} -U(\chi),\label{n}
\ee
where $n=1,2,3...$; although $n=1$ reproduces the standard model, the other values of $n$ define many other generalized models. The first model is the well-known $\phi^4$ model, with spontaneous symmetry breaking; so, it has the kinklike solution $\phi(x)=\tanh(x)$. We use the above procedure to get to the first-order equation $\phi^\prime=1-\phi^2$, which has the solution $\phi(x)=\tanh(x)$, as expected. In the second model \eqref{n} we leave $U(\chi)$ arbitrary. We now take as deformation the identity function, $g(\chi)=\chi$, and we then obtain $\chi^\prime=1-\chi^2$, which also has the same solution, $\chi(x)=\tanh(x)$. However, for the model \eqref{n} the potential has to have the form
\be\label{Un}
U(\chi)=\frac{2n-1}{2n}(1-\chi^2)^{2n}.
\ee
The result is remarkable: it introduces a family of generalized models, with $n=2,3,4,...,$ all having the very same static structure.

In order to further illustrate the procedure, let us consider as the starting Lagrange density, a model engendering polynomial or nonpolynomial potential.
In the case of polynomial potential we take the model 
\be
{\cal L}=-X^{2} -\frac{3}{4}(1-\phi^2)^{2}.
\ee
Its static solution has the form
 \begin{eqnarray}
\phi(x) &=& \left\{
\begin{array}{ll}
-1; &  \mbox{ for } x<-{\pi}/{2},\\ 
\sin(x); & \mbox{ for } -{\pi}/{2}\leq x\leq {\pi}/{2}, \\ 
1; & \mbox{ for } x>{\pi}/{2}.
\end{array} \right.
\end{eqnarray}
It is a compacton, a localized structure that lives in a compact space; its energy density vanishes outside the interval $[-\pi/2, \pi/2]$. The first-order equation is $\phi^\prime=(1-\phi^2)^{1/2}$. The model also supports anticompactons, but we leave this aside in this work.

\begin{figure}[ht]
\begin{center}
\includegraphics[{height=4cm,width=8cm,angle=-00}]{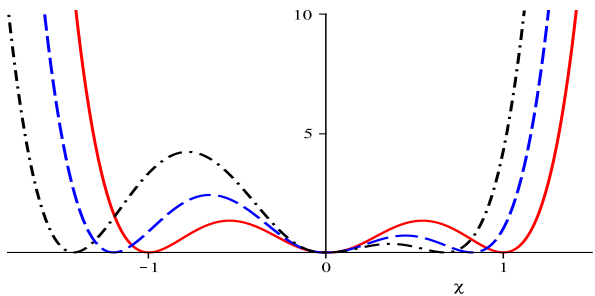}
\end{center}
\caption{(Color online) The potential \eqref{u1} with $n=2$ and $p=3$, for $a=0$ (red, solid line),  $a=1/16$ (blue, dashed line), and $a=1/8$ (black, dotted-dashed line).}\label{fig1}
\end{figure}

We take as the second model, the model \eqref{n}, and we illustrate this case considering two distinct functions: $g_1(\chi)=\pm(a+\chi^{1/p})$ and $g_2(\chi)=\pm(1-g_1^2)^{1/2}$, with $p$ being an odd integer, $p>1$ and $a\in (0,1].$ Interestingly, we see that both functions $g_1$ and $g_2$ lead to the same $S(\chi)$, which has the form
\be
S(\chi)=p\,\phi^{1-1/p}\left(1-(a+\chi^{1/p})^2\right)^{1/2}\,.
\ee
The potential is then given by 
\be\label{u1}
U(\chi)=\frac{(2n\!-\!1)p^{2n}}{2n}\chi^{2n-2n/p}\!\left(1\!-\!(a\!+\!\chi^{1/p})^2\right)^n.
\ee
It has three minima, at ${\bar\chi}_0=0$ and ${\bar\chi_\pm}=\pm(1\mp a)^p$. It is depicted in Fig.~\eqref{fig1}, for $p=3$, and for $a=0,1/16$ and $1/8$.

\begin{figure}[ht]
\begin{center}
\includegraphics[{width=8cm,angle=-00}]{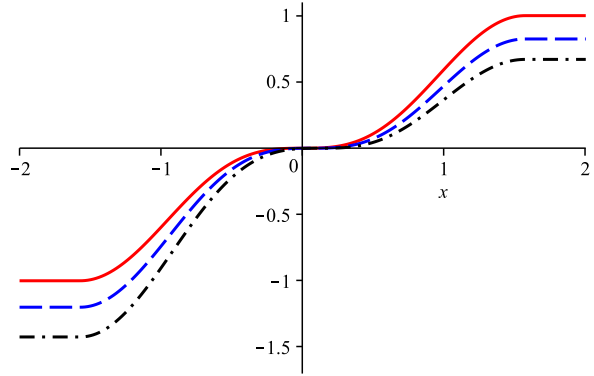}
\end{center}
\caption{(Color online) The double compacton \eqref{dc}, depicted for $p=3$ and $a=0$ (red, solid line),  $a=1/16$ (blue, dashed line), and $a=1/8$ (black, dotted-dashed line).}\label{fig2}
\end{figure}

The static solutions are also compactons;  if we use the inverse of $g_1$, we can connect the minima ${\bar\chi}_\pm$; the solutions have the form
\begin{eqnarray}\label{dc}
\chi(x) &=& \left\{
\begin{array}{ll}
-(1+ a)^p; \;\;x<-{\pi}/{2},\\ 
-(a-\sin(x))^p; \;\; -{\pi}/{2}\leq x\leq {\pi}/{2},\;\;\;\;\;\;\\ 
(1- a)^p; \;\; x>{\pi}/{2}.
\end{array} \right.
\end{eqnarray}
In Fig.~\eqref{fig2} we depict some solutions, for specific values of the parameters $a$ and $p$. These compactons have the 2-compact profile, similar to the 2-kink form found in \cite{bmm}. As far as we know, it is the first time a compact structure like this appears in physics.

\begin{figure}[ht]
\begin{center}
\includegraphics[{height=4cm,width=8cm,angle=-00}]{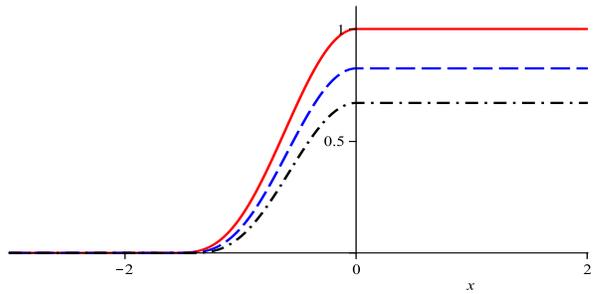}
\end{center}
\caption{(Color online) The compact solution \eqref{c2},  depicted for $p=3$ and $a=0$ (red, solid line),  $a=1/16$ (blue, dashed line), and $a=1/8$ (black, dotted-dashed line).}\label{fig3}
\end{figure}

We now use the inverse of $g_2(\chi)$, and make the solutions to connect the minima ${\bar\chi}_-$ and ${\bar\chi}_0$; the compact solutions are
\begin{eqnarray}\label{c1}
\chi(x) &=& \left\{
\begin{array}{ll}
0; \;\; x\geq{\pi}/{2},\\ 
-(a+ \cos(x))^p; \;\; 0\leq x\leq {\pi}/{2}, \\ 
-(1+a)^p; \;\; x\leq0.
\end{array} \right.
\end{eqnarray}
 We can also use the inverse of $g_2$ to connect the minima $\bar{\chi}_0=0$ and $\bar{\chi}_+$; the compact solutions are
 \begin{eqnarray}\label{c2}
\chi(x) &=& \left\{
\begin{array}{ll}
(1-a)^p;\;\; x\geq 0,\\ 
-(a-\cos(x))^p; \;\; -{\pi}/{2}\leq x\leq 0, \\ 
0; \;\; x\leq -{\pi}/{2}.
\end{array} \right.
\end{eqnarray}
 We illustrate the compact solutions \eqref{c2} in Fig.~\eqref{fig3}.

Let us now take as the starting Lagrange density another model, described by nonpolynomial potential,
\be
{\cal L}=-X^2-\frac34(1-\cos(\phi))\,.
\ee
It is generalized sine-Gordon model, and has the compact solutions
\begin{eqnarray}\label{phi61} 
\phi(x)&\!\!\!=\!\!\!&\left\{
\begin{array}{lll}
2m\pi;\;\;x\leq -3{\bar x},\\
f(x)+2m\pi;-3{\bar x}\leq x\leq -{\bar x},\;\;\;\\
2(m+1)\pi;\;\;\;x\geq -{\bar x}.
\end{array}\right.
\end{eqnarray}
where  $m=0, \pm1, \pm2,...$,
\be
f(x)=4\,\arctan Z^2(x),
\ee
and 
\be
Z(x)=\frac{{\rm sn}(bx,k)-(1+\sqrt{2}){\rm cn}(bx,k)}{{\rm sn}(bx,k)+(1+\sqrt{2}){\rm cn}(bx,k)},
\ee
and
\be
k^2=4(3\sqrt{2}-4),\;\;\;b^2=\frac1{64}(3\sqrt{2}+4).
\ee
Also, ${\bar x}=4.409757$, and $\rm {sn}$ and $\rm {cn}$ are Jacobi elliptic functions.


We consider the functions
\be\label{def6}
g_{\pm}(\chi)=\pm\chi\pm \alpha\sin^2(s\chi)\,,
\ee
where $s$ is real number and $\alpha$ is small, $\alpha\approx 0$. We follow the work before the last one in Ref.~\cite{deform} and, up to first-order in $\alpha$, we get the potential
\bes\label{sGp}\ben
U(\chi)=\frac{(2n-1)}{2n}{S}^{2n}(\chi)\,,
\een
where
\ben\label{wc1}
{S}^2(\chi)&=&\sqrt{2}\sin\left(\frac{\chi}{2}\right)+\frac{\alpha}{\sqrt{2}}\bigg(\cos\left(\frac{\chi}{2}\right)\sin^2\left({s\chi}\right)\nonumber\\
&&-4s\sin\left(\frac{\chi}{2}\right)\sin\left({2s\chi}\right)\bigg)\,.
\een\ees
Also, taking the inverse of \eqref{def6} gives 
\be
g^{-1}_{\pm}(\phi)=\pm \phi\mp\alpha\sin^2(s\phi)\,,
\ee
where $\phi=\phi(x)$ is given by \eqref{phi61}.

\begin{figure}[t]
\includegraphics[{width=8cm,angle=-00}]{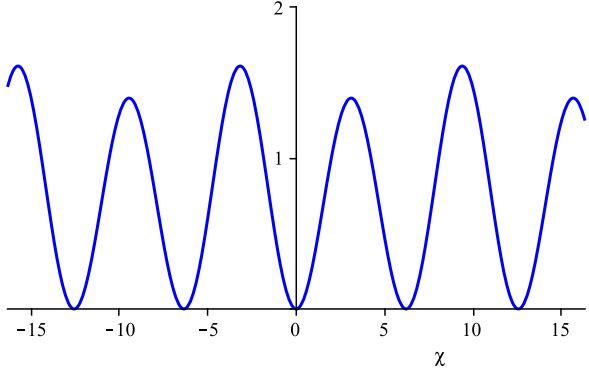}
\caption{The potential \eqref{sGp} for $n=2$, and for $\alpha=0.1$ and $s=-1/4$, showing the two, large and small sectors.}\label{fig4}
\end{figure}

\begin{figure}[ht]
\includegraphics[{height=2.6cm,width=8cm,angle=-00}]{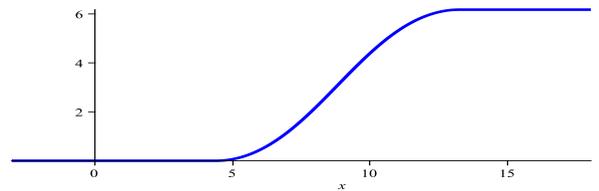}
\caption{The small compact solution \eqref{sc}.}\label{fig5}
\end{figure}

The parameter $s$ determines the multiplicity of distinct topological sectors of the generalized sine-Gordon potential \eqref{sGp}. For instance, taking $s=-1/4$ or $s=-1/3$ leads us to the double or triple sine-Gordon model, respectively. We illustrate this with the double sine-Gordon model, which has the minima  $\bar\chi_{\pm}=\pm 2\pi-\alpha$ next to the minimum $\chi=0$; see Fig.\eqref{fig4}. For $\bar\chi_{-}\leq\chi\leq\ 0$ we have the large compact solution
\begin{eqnarray}\label{lc}
\chi(x) &=& \left\{
\begin{array}{ll}
0; \;\;\; x\geq {\bar x},\\ 
-f(x)\!-\!\alpha\sin^2(f(x)/4);\;-{\bar x}\leq x\leq {\bar x},\;\;\;\\
\bar\chi_{-};\;\; x\leq -{\bar x}.\;
\end{array} \right.
\end{eqnarray}
Also, for $0\leq\chi\leq\bar\chi_+$, the small compact solution is
\begin{eqnarray}\label{sc}
\chi(x) &=& \left\{
\begin{array}{ll}
\bar\chi_{+}; \;\;\; x\geq 3{\bar x},\\ 
f(x)\!-\!\alpha\sin^2(f(x)/4);\;{\bar x}\leq x\leq 3{\bar x},\;\;\;\\
0;\;\; x\leq {\bar x}.
\end{array} \right.
\end{eqnarray}
In Fig.~\eqref{fig5}, we depict the small compact solution of the double sine-Gordon potential displayed in Fig.~\eqref{fig4}. A family of new models appears, controlled by $s$.

{\it Stability.} We examine linear stability for  the generalized model
\be\label{gm}
{\cal L}=(-1)^{n-1}\frac{2^{n-1}}{n}X^{n} -V(\phi),
 \ee
for $n=1,2,...\,$. As we have shown, it may support static solution $\phi(x)$ such that $\phi^\prime=R(\phi)$, and now we consider $\phi(x,t)=\phi(x)+\eta(x,t)$, where $\eta(x,t)=\eta(x)\cos(wt)$ stands for small fluctuation around the static solution. Up to first-order in the fluctuation, we obtain
\be\label{sch}
\left(-\frac{d^2}{dx^2}+U(x)\right)u(x)=\frac{\omega^2}{2n-1} u(x)\,,
\ee
where 
\be
U(x)=\frac{((F_X)^{1/2})_{xx}}{(F_X)^{1/2} }+\frac{V_{\phi\phi}}{F_X },
\ee
with $u\!=\!( F_X)^{1/2}\eta$. We use Eq.~\eqref{gm} to get
$U=n^2 R^2_\phi+nRR_{\phi\phi}$, and the above Schroedinger-like equation can be factorized as
\be
\left(\frac{d}{dx}+nR_\phi\right)\left(-\frac{d}{dx}+nR_\phi\right)u(x)=\frac{w^2}{2n-1} u(x).
\ee
The Hamiltonian is non-negative, and so there is no negative eigenvalue. The static solution is stable, and the zero mode obeys $u_0(x)\!\sim\! R^{n}(\phi(x))$.

{\it Ending comments.} In this work we developed a method to construct and solve generalized models described by a single real scalar field in (1,1) spacetime dimensions.
The approach is simple and direct, leading us to new models together with their respective localized structures. We focused mainly on solutions with compact profile, showing
how to obtain compact structures from generalized models with distinct potentials, engendering polynomial or nonpolynomial interactions. We also investigated linear stability,
showing how it works explicitly. The results show that the compact solutions we have found are all stable.

The procedure suggested in this work is robust and can be used in a diversity of ways, to help us explore new models and the classical structures they may engender.
The method works with ordinary differential equations, so we can think of using it to describe more complex static structures in higher spacetime dimensions. The problem here is that localized, spherically symmetric topological solutions such as vortices and monopoles, for instance, in general require the presence of gauge fields, Abelian and non Abelian, respectively, and this makes the problem much harder. We shall further report on this elsewhere.

{\it Acknowledgments.} The authors thank CAPES and CNPq for partial financial support.



\begin{thebibliography}{99}


\bb{soliton}
M. Ablowitz and P.  Clarkson, 
{\it Solitons, nonlinear evolution equations and inverse scattering} (Cambridge, Cambridge/UK, 1991).

\bb{compact1}
P. Rosenau and J.M. Hyman, Phys. Rev. Lett. {\bf70}, 564 (1993).

\bb{compact2}
P. Rosenau, Phys. Rev. Lett. {\bf73}, 1737 (1994);
P. Rosenau and A. Pikovsky, Phys. Rev. Lett. 94, 174102 (2005);
P. Rosenau, J.M. Hyman, and M. Staley, Phys. Rev. Lett. {\bf98}, 024101 (2007);
P. Rosenau and E. Kashdan, Phys. Rev. Lett. {\bf101}, 264101 (2008);
P. Rosenau and E. Kashdan, Phys. Rev. Lett. {\bf104}, 034101 (2010);
F.Kh. Abdullaev, P.G. Kevrekidis, and M. Salerno, Phys. Rev. Lett. 105, 113901 (2010).

\bb{css}
F. Cooper, H. Sheppard, and P. Sodano, Phys. Rev. E {\bf48}, 4027 (1993);
A. Khare and F. Cooper, Phys. Rev. E {\bf48}, 4843 (1993);
F. Cooper, J.M. Hyman, and A. Khare, Phys. Rev. E {\bf64}, 026608 (2001).

\bibitem{book}
N. Manton and P. Sutcliffe, {\it Topological solitons} (Cambridge, Cambridge/UK, 2004).

\bb{kink}
M. Toharia and M. Trodden, Phys. Rev. Lett. {\bf100}, 041602 (2008);
G. Basar and G.V. Dunne, Phys. Rev. Lett. {\bf100}, 200404 (2008);
J. Belmonte-Beitia, V.M. Perez-Garcia, V. Vekslerchik, and V.V. Konotop, Phys. Rev. Lett. {\bf100}, 164102 (2008);
A.T. Avelar, D. Bazeia, and W.B. Cardoso, Phys. Rev. E {\bf79}, 025602(R) (2009);
A. Vanhaverbeke, A. Bischof, and R. Allenspach, Phys. Rev. Lett. {\bf101}, 107202 (2008);
A. Alonso-Izquierdo, M.A. Gonzalez Leon, and J. Mateos Guilarte, Phys. Rev. Lett. {\bf101}, 131602 (2008);
R. Auzzi and S. Prem Kumar, Phys. Rev. Lett. {\bf103}, 231601 (2009);
A. D. Martin and J. Ruostekoski, Phys. Rev. Lett. {\bf104}, 194102 (2010);
T. Romanczukiewicz and Ya. Shnir, Phys. Rev. Lett. {\bf105}, 081601 (2010);
M. Angeles Perez-Garcia, J. Silk, and J. R. Stone, Phys. Rev. Lett. {\bf105}, 141101 (2010).

\bb{RS}
L. Randall and R. Sundrum, Phys. Rev. Lett. {\bf83}, 4690 (1999);
W.D. Goldberger and M.B. Wise, Phys. Rev. Lett. {\bf83}, 4922 (1999);
O. DeWolfe, D.Z. Freedman, S.S. Gubser, A. Karch, Phys. Rev. D {\bf62}, 046008 (2000).

\bb{brane1}
A. Campos, Phys. Rev. Lett. {\bf88}, 141602 (2002);
D. Bazeia, C. Furtado, and A.R. Gomes, JCAP {\bf 0402}, 002 (2004);
D. Bazeia and A.R. Gomes, JHEP {\bf0405}, 012 (2004).

\bb{defect}
E. Babichev, Phys. Rev. D {\bf74}, 085004 (2006);
D. Bazeia, L. Losano, R. Menezes, and J. C. R. Oliveira, Eur. Phys. J. C {\bf51}, 953 (2007);
C. Adam, J. Sanchez-Guillen, and A. Wereszczynski, J. Phys. A {\bf40}, 13625 (2007); Erratum-ibid. A {\bf42}, 089801 (2009);
M. Olechowski, Phys. Rev. D {\bf78}, 084036 (2008);
C. Adam, N. Grandi, J. Sanchez-Guillen, and A. Wereszczynski, J. Phys. A {\bf41}, 212004 (2008);
C. Adam, N. Grandi, P. Klimas, J. Sanchez-Guillen, and A. Wereszczynski, J. Phys. A {\bf41}, 375401 (2008);
C. Adam, P. Klimas, J. Sanchez-Guillen, and A. Wereszczynski, J. Phys. A {\bf42}, 135401 (2009);
D. Bazeia, A. R. Gomes, L. Losano, and R. Menezes, Phys. Lett. B {\bf671}, 402 (2009);
C. Adam, P. Klimas, J. Sanchez-Guillen, and A. Wereszczynski, J. Math. Phys. {\bf50}, 102303 (2009);
D. Bazeia, E. da Hora, R. Menezes, H.P. de Oliveira, and C. dos Santos, Phys. Rev. D {\bf81}, 125016 (2010);
C. Adam, N. Grandi, P. Klimas, J. Sanchez-Guillen, and A. Wereszczynski, Gen. Rel. Grav. {\bf42}, 2663 (2010);
C.A.G. Almeida, D. Bazeia, L. Losano, and R. Menezes, Phys. Rev. D {\bf88}, 025007 (2013);
D. Bazeia, A.S. Lobao Jr., L. Losano, and R. Menezes, Phys. Rev. D {\bf88}, 045001 (2013). 

\bb{optics}
B.A. Malomed, {\it Soliton management in periodic systems} (Springer-Verlag, Berlin, 2007). 

\bibitem{first}
D. Bazeia,  L. Losano, and R. Menezes, Phys. Lett. B {\bf 668}, 246 (2008).

\bb{blm}
D. Bazeia, L. Losano, and J.M.C Malbouisson, Phys. Rev. D {\bf66}, 101701(R) (2002).

\bb{kf}
C. Armendariz-Picon, T. Damour, and V. Mukhanov, Phys. Lett. B {\bf458}, 209 (1999);
C. Armendariz-Picon, V. Mukhanov, and P.J. Steinhardt, Phys. Rev. Lett. {\bf85}, 4438 (2000);
C. Armendariz-Picon, V. Mukhanov, and P.J. Steinhardt, Phys. Rev. D {\bf63}, 103510 (2001).

\bb{twin}
M. Andrews, M. Lewandowski, M. Trodden, and D. Wesley, Phys. Rev. D {\bf82}, 105006 (2010);
P.P. Avelino, D. Bazeia, R. Menezes, and J.G.G.S. Ramos, Eur. Phys. J. C {\bf71}, 1683 (2011); 
D. Bazeia, J.D. Dantas, A.R. Gomes, L. Losano, and R. Menezes, Phys. Rev. D {\bf84}, 045010 (2011);
C. Adam and J.M. Queiruga, Phys. Rev. D {\bf84}, 105028 (2011);
D. Bazeia and R. Menezes, Phys. Rev. D {\bf84}, 125018 (2011);
C. Adam and J.M. Queiruga, Phys. Rev. D {\bf85}, 025019 (2012);
D. Bazeia, E. da Hora, and R. Menezes, Phys. Rev. D {\bf85}, 045005 (2012);
D. Bazeia and J. D. Dantas, Phys. Rev. D {\bf85}, 067303 (2012);
D. Bazeia, A. S. Lob\~ao Jr, and R. Menezes, Phys. Rev. D {\bf86}, 125021 (2012).



\bb{deform}
C.A.G. Almeida, D. Bazeia, L. Losano, and J.M.C. Malbouisson, Phys. Rev. D {\bf69}, 067702 (2004);
A. de Souza Dutra and A.C. Amaro de Farias Jr., Phys. Rev. D {\bf72}, 087701 (2005);
D. Bazeia and L. Losano, Phys. Rev. D {\bf73}, 025016 (2006); D. Bazeia, M.A. Golzalez Leon, L. Losano, and J. Mateos Guilarte, Phys. Rev. D {\bf73}, 105008 (2006);
D. Bazeia, Ashok Das, L. Losano, and A. Silva, Ann. Phys. {\bf 323}, 1150 (2008);
V.I. Afonso, D. Bazeia, M.A. Gonzalez Leon, L. Losano, and J. Mateos Guilarte, Nucl. Phys. B {\bf810}, 427 (2009);
A.E.R. Chumbes and M.B. Hott, Phys. Rev. D {\bf81}, 045008 (2010);
D. Bazeia, Ashok Das, L. Losano, and M.J. Santos, Appl. Math. Lett. {\bf23}, 681 (2010);
A.E. Bernardini and R. da Rocha, Adv. High Energy Phys. {\bf2013}, 304980 (2013);
D. Bazeia, L. Losano, R. Menezes, and R. da Rocha, Eur. Phys. J. C {\bf73}, 2499 (2013); 
C. Adam, L. A. Ferreira, E. da Hora, A. Wereszczynski, and W. J. Zakrzewski, JHEP {\bf1308}, 062 (2013).

\bb{bmm}D. Bazeia, J. Menezes, and R. Menezes, Phys. Rev. Lett. {\bf91}, 241601 (2003).

\end{thebibliography}
\end{document}